\documentclass[12pt]{article}
\usepackage{a4wide}
\usepackage{graphicx}
\newcommand{\be}{\begin{equation}}
\newcommand{\ee}{\end{equation}}
\newcommand{\ba}{\begin{eqnarray}}
\newcommand{\ea}{\end{eqnarray}}

\begin{document}

\begin{titlepage}
\begin{flushright}
\end{flushright}
\vfill
\begin{center}
{\Large {\bf Pion mass dependence of the $K_{l3}$ semileptonic scalar form factor within finite volume} }
\vfill
{\bf K. Ghorbani$^{ a,d}$, M. M. Yazdanpanah$^{ b,d}$ and A. Mirjalili$^{ c,d} $ }\\[1cm]
\end{center}
{a) Physics Department, Faculty of Sciences, Arak University, Arak 38156-8-8349, Iran\\
 b) Physics Department, Kerman Shahid Bahonar University, Kerman, Iran\\
 c) Physics Department, Yazd University, 89195- 741, Yazd, Iran \\
 d) School of Particles and Accelerators, IPM(Institute for research in fundamental sciences), P.O.Box 19395-5531, Tehran, Iran
}
\vfill
\begin{abstract}
We calculate the scalar semileptonic kaon decay in
finite volume at the momentum transfer $t_{m} = (m_{K} - m_{\pi})^2$, using
chiral perturbation theory. At first we obtain the hadronic
matrix element to be calculated in finite volume.
We then evaluate the finite size effects for two volumes
with $L = 1.83~fm$ and $L= 2.73~fm$ and find that the
difference between the finite volume corrections
of the two volumes are larger than
the difference as quoted in \cite{Boyle2007a}.
It appears then that the pion masses used for the scalar form factor
in ChPT are large which result in large finite volume corrections.
If appropriate values for pion mass are used, we believe that
the finite size effects estimated in this paper can be useful for
Lattice data to extrapolate at large lattice size.
\end{abstract}
\vfill
\vfill

\end{titlepage}

\section{Introduction}
\label{intro}
A precise determination of the CKM matrix element, $V_{us}$, is highly demanded if
one expects to find a footprint of new physics in the unitarity relation of the first
row of the CKM matrix \cite{Cabibbo:1963,Kobayashi:1973}, namely,
\begin{equation}
\label{CKM}
|V_{ud}|^2+|V_{us}|^2+|V_{ub}|^2 = 1.
\end{equation}
Among three entities above, $|V_{us}|$ is less precisely known from experimental
measurements in strangeness changing weak kaon decays $K_{l3}$ and therefore dominates
the uncertainty in the unitarity relation. The theoretical framework for the determination
of this quantity is explained by Leutwyler and Roos in \cite{Leutwyler:1984}.
To achieve a more precise value of $|V_{us}|$, one requires a better control over
the long distance effects due to the strong interactions in the process under consideration.
In fact one can only obtain experimentally, the combination $|V_{us}f_{+}(0)|$ by measuring
the semileptonic kaon decay rates. Here, $f_{+}(0)$ denotes the related vector form factor at
zero momentum transfer.
This can be exemplified in the master formula for $K_{l3}$,
see \cite{Cirigliano:2007zz} and references therein
\begin{eqnarray}
\Gamma\left(K^i\to\pi^j\ell^+\nu_l\right)
= C^{ij^2}\frac{G_F^2 S_{EW} m_K^5}{192\pi^3}
\left|V_{us} f_+^{K^i\pi^j(0)}\right|
{\cal I}^{ij}_\ell
\left(1+2\Delta_{EM}^{ij}\right)\,,
\end{eqnarray}
where $G_{F}$ is the fermi constant, $S_{EW}$ incorporates the short distance
electroweak corrections and ${\cal I}^{ij}_\ell$ is the phase-space integral depending
on the slope and curvature of the form factors.
$f_{+}(0)$ contains the non-perturbative effects of strong interactions
whose precision estimate may lower the uncertainties on $|V_{us}|$.
There exist two main approaches for the calculation of the vector form factors.
One powerful technique is the application of chiral perturbation theory (ChPT)
as an effective field theory describing low energy strong interactions.
The determination of $f_{+}(0)$ at one loop order within ChPT was evaluated
by Gasser and Leutwyler \cite{Gasser-Leutwyler}. Along the same lines the
calculations were extended further to incorporate two loop order corrections \cite{Bijnens-Talavera}.
Despite its model-independent prediction, however, ChPT calculations
may become less predictive at higher orders due to undetermined parameters.

Lattice QCD on the other hand, is a brute force approach to determine
the QCD observables by evaluating the functional integral of QCD numerically.
In recent years the technical procedures have reached to a point that
calculations with relatively small quark masses are feasible,
see \cite{Bazavov:2009fk,Saoki2009,Saoki2010,DurrScience,Durr2010}
and references therein.
Lattice data are not free from systematic errors e.g.
finite volume effects and chiral extrapolation.
As soon as pion mass is small enough and lattice size is large enough in order
to make legitimate application of ChPT for the extrapolations, one can get reliable
predictions out of Lattice QCD data.
There are a large amount of works in the
literature on exploiting ChPT to correct the systematic errors.
Lattice size effects on pion mass are studied in \cite{Colangelo2006}
and for pion decay constant in \cite{Colangelo2005,Colangelo2004}.
Along the same arguments this type of calculations are done for quark vacuum
expectation values \cite{Bijnens-Ghorbani2006}.
Moreover, finite volume effects on pion pion scattering parameters are evaluated
near threshold \cite{Bedaque2006}.
There exist Lattice data on the semileptonic vector and scalar form factors, see
for example \cite{Becirevic2004,Boyle2007a,Boyle2007b,Lubicz,Juttner}.
As suggested in \cite{Becirevic2004,Hashimoto1999} the starting point in doing these
calculations in the Lattice is to obtain the scalar form factor at the maximum momentum
transfer $t_{m} = (m_{K}-m_{\pi})^2$, where high statistical
accuracy is achievable. We then deem interesting to evaluate the
finite size effects for the scalar semileptonic kaon decay. This decay
gives the important key role in kaon physics.
Therefore we set the main motivation of the present work as pushing the line
of the ChPT applications to a case which involves external momentum. For the sake of
simplicity we assume external momentum takes on zero spatial component.
To this end, we have checked how far we can get on extending
the reliability of ChPT to doing chiral and finite volume extrapolations.
Earlier work on finite volume calculations for meson current matrix element with
external spatial momentum can be found in \cite{Bunton}.

The outline of this article is as follows. We begin with an introduction
about ChPT in finite volume and infinite volume in Sec.\,\ref{chpt}.
On Sec.\,\ref{kl3defin} the definition of the $K_{l3}$ form factors
are reviewed and in Sec.\,\ref{loopresult} one loop result of the
hadronic matrix element in infinite volume and in a tensor form is provided.
Finite volume calculations entailing necessary Feynman integrals
in finite volume are sketched in Sec.\,\ref{finiteresult}.
The last section reports on the finite size effects on the scalar form factor
and ends with conclusion.

\section{ChPT in infinite and finite volume}
\label{chpt}
\subsection{ChPT in infinite volume}
Chiral perturbation theory (ChPT) is an effective field theory which describes
the low energy dynamics of Quantum Chromo Dynamics (QCD).
Weinberg in a seminal paper for the first time paved the way for a systematic
application of effective Lagrangian \cite{Weinberg0}.
Pseudo-Goldston mesons play the role of dynamical degrees of freedom in the effective Lagrangian.
These mesons are resulted as a consequence of the spontaneous breakdown of the chiral symmetry in QCD.
The expansion parameter is generically in terms of the momentum, $p^2$, and quark mass, $m_{q}$.
Later on Gasser and Leutwyler in two elegant papers extended the Lagrangian to order $p^4$ \cite{GL0,GL1}.
One can arrange the full effective Lagrangian as series of operators with increasing
number of derivatives and increasing power of quark mass
\begin{equation}
\label{lagL2}
\mathcal{L}_{eff} = \mathcal{L}_{2} + \mathcal{L}_{4} + \mathcal{L}_{6}
 + \cdots
\,.
\end{equation}
Here the subscripts show the chiral order.
The lowest order Lagrangian involving two terms has the following form
\begin{equation}
\mathcal{L}_{2} = \frac{F_0^{2}}{4} \left( \langle D_{\mu}U D^{\mu}U^{\dagger}
 \rangle +
\langle\chi U^{\dagger} + U \chi^{\dagger}  \rangle  \right).
\end{equation}
The mass relation $m_\pi^2 = B_0\left(m_u+m_d\right)$ allows one to count
quark masses as order $p^{2}$.
The next to leading order Lagrangian consists of 10+2 independent operators
with corresponding effective low energy constants (LEC's)
\begin{eqnarray}
\label{lagL4}
\mathcal{L}_{4} &&\hspace{-0.1cm} =
L_1 \langle D_\mu U^\dagger D^\mu U \rangle^2
+L_2 \langle D_\mu U^\dagger D_\nu U \rangle
     \langle D^\mu U^\dagger D^\nu U \rangle \nonumber\\&&\hspace{-0.1cm}
+L_3 \langle D^\mu U^\dagger D_\mu U D^\nu U^\dagger D_\nu U\rangle
\nonumber\\&&\hspace{-0.1cm}
+L_4 \langle D^\mu U^\dagger D_\mu U \rangle
 \langle \chi^\dagger U+\chi U^\dagger \rangle
\nonumber\\&&\hspace{-0.1cm}
+L_5 \langle D^\mu U^\dagger D_\mu U (\chi^\dagger U+U^\dagger \chi ) \rangle
+L_6 \langle \chi^\dagger U+\chi U^\dagger \rangle^2
\nonumber\\&&\hspace{-0.1cm}
+L_7 \langle \chi^\dagger U-\chi U^\dagger \rangle^2
+L_8 \langle \chi^\dagger U \chi^\dagger U
 + \chi U^\dagger \chi U^\dagger \rangle
\nonumber\\&&\hspace{-0.1cm}
-i L_9 \langle F^R_{\mu\nu} D^\mu U D^\nu U^\dagger +
               F^L_{\mu\nu} D^\mu U^\dagger D^\nu U \rangle
\nonumber\\&&\hspace{-0.1cm}
+L_{10} \langle U^\dagger  F^R_{\mu\nu} U F^{L\mu\nu} \rangle
+H_1 \langle F^R_{\mu\nu} F^{R\mu\nu} + F^L_{\mu\nu} F^{L\mu\nu} \rangle
\nonumber\\&&\hspace{-0.1cm}
+H_2 \langle \chi^\dagger \chi \rangle\,.
\end{eqnarray}
The notation $\langle...\rangle$ = $ \mathrm{Tr}_F\left(...\right)$ indicates the trace over the flavors.
The matrix $U \in SU(3)$ parameterizes the eight light mesons in an exponential representation
\begin{equation}
U(\phi) = \exp(i \sqrt{2} \phi/F_0)\,,
\end{equation}
where
\begin{eqnarray}
\phi (x)
 = \, \left( \begin{array}{ccc}
\displaystyle\frac{ \pi_3}{ \sqrt 2} \, + \, \frac{ \eta_8}{ \sqrt 6}
 & \pi^+ & K^+ \\
\pi^- &\displaystyle - \frac{\pi_3}{\sqrt 2} \, + \, \frac{ \eta_8}
{\sqrt 6}    & K^0 \\
K^- & \bar K^0 &\displaystyle - \frac{ 2 \, \eta_8}{\sqrt 6}
\end{array}  \right) .
\end{eqnarray}
The covariant derivative and the field strength tensor
are defined as
\begin{equation}
D_\mu U = \partial_\mu U -i r_\mu U + i U l_\mu \,, \quad
F_{\mu\nu}^{L} = \partial_\mu l_\nu -\partial_\nu l_\mu
-i \left[ l_\mu , l_\nu \right]\,,
\end{equation}
and a similar definition apply to the right-handed field strength.
Here $l_\mu$ and $r_\mu$ represent the left-handed and
right-handed chiral currents respectively.
$\chi$ is parameterized in terms of scalar ($s$) and pseudo scalar ($p$)
external densities as $\chi = 2B_0 \left( s + i p\right)$.
For the process discussed in this article it is sufficient to set

\begin{eqnarray}
s=\left( \begin{array}{ccc}
\displaystyle \hat m & & \\
 &\displaystyle  \hat m  &  \\
 & &\displaystyle m_{s}
\end{array}  \right)
\,, l_{\mu} = \frac{g_{2}}{\sqrt{2}} \left( \begin{array}{ccc}
\displaystyle  & V_{ud} W_{\mu}^{+} & V_{us} W_{\mu}^{+} \\
V_{ud}^{\ast} W_{\mu}^{-}   &  &  \\
V_{us}^{\ast} W_{\mu}^{-}   &  &
\end{array}\right),
\nonumber\\&&\hspace{-9cm}
r_{\mu} = 0.
\nonumber\\
\end{eqnarray}
The weak coupling constant, $g_{2}$, is related to the Fermi constant by
\begin{equation}
\frac{g_{2}^2}{8m_{W}^2} = \frac{G_{F}}{\sqrt{2}}.
\end{equation}
\subsection{ChPT in finite volume}

The effective Lagrangian approach is also applicable to a system which is bound
to a large space of volume $V = L^3$ with an infinite time extent.
The effective Lagrangian to be used for a system enclosed into a finite size
is the same as one we use for processes which occur in infinite volume with the same
values for the low energy effective constants.
The pioneering works \cite{Gasser1986finite1,Gasser1987finite2,Gasser1987finite3}
provide the foundations in this regards.
To do the calculations in finite volume as suggested in \cite{Gasser1987finite3}
one imposes a
periodic boundary condition on particle fields which results in the momentum
quantization and consequently the modifications of the quantum corrections.
The propagation of a particle in space-time is defined through the two-point
correlation function. Therefore the correlation function in finite volume becomes
\begin{equation}
G_{V} = \frac{1}{L^3} \sum_{\vec{q}} \int \frac{dq^0}{2\pi} G(q^0,\vec{q}),
\end{equation}
where the three dimensional momentum takes the discrete values
\begin{equation}
\vec{q} = \frac{2\pi}{L} \vec{n},
\end{equation}
with $\vec{n}$ is a three dimensional vector with integral components and $L$ is
the linear size of the box.
Since the reliability of ChPT is limited to the small momenta, in finite volume
this leads to the condition that
\begin{equation}
F_{\pi} L >>1 \,,
\end{equation}
where $F_{\pi}$ is the pion decay constant.
In addition we stay in a limit where the Compton wavelength of the pion is smaller than
the lattice size, $L$, so that the zero mode of the pion field does not become strongly coupled.
This condition is satisfied if
\begin{equation}
m_{\pi} L >>1 \,.
\end{equation}
This is the so-called {\it p-regime} in which we do our calculations here.
Moreover, the quantity $m_{\pi}L$ plays the role of the power counting
in the perturbative calculations.

\section{The definition of the $K_{l3}$ form factors}
\label{kl3defin}
The semileptonic kaon weak decays which are generally shown as $K_{l3}$ are:
\begin{equation}
K^{+}(p) \to \pi^{0}(p^{\prime}) l^{+}(p_{l}) \nu_{l}(p_{\nu}) \,,
\end{equation}
\begin{equation}
K^{0}(p) \to \pi^{-}(p^{\prime}) l^{+}(p_{l}) \nu_{l}(p_{\nu}).
\end{equation}
Here $l$ stands for $electron$ and $muon$. The two other processes are charge conjugate modes of the processes above. \\
Neglecting scalar and tensor contributions, the matrix element takes on the structure
\begin{equation}
{\cal M} = \frac{G_{F}}{\sqrt{2}} V^{\ast}_{us} J^{\mu} F^{+}_{\mu} (p^{\prime},p),
\end{equation}
with \\
\begin{eqnarray}
J^{\mu} &&\hspace{-0.1cm} = {\bar u}(p_{\nu}) \gamma^{\mu}(1-\gamma_{5}) v(p_{l}),
\nonumber\\
F^{+}_{\mu}(p^{\prime},p) &&\hspace{-0.1cm}= \hspace{0.1cm}<\pi^{0}(p^{\prime})| {\bar s} \gamma_{\mu} u(0)|K^{+}(p)>.
\end{eqnarray}
The general form of the vector matrix element is defined
\begin{eqnarray}
<\pi^{0}(p^{\prime})| {\bar s} \gamma_{\mu} u(0)|K^{+}(p) &&\hspace{-0.1cm}=
\nonumber\\&&\hspace{-2cm}
\frac{1}{\sqrt{2}}[ (p+p^{\prime})_{\mu} f^{K^{+}\pi^{0}}_{+}(t)
+(p-p^{\prime})_{\mu}f^{K^{+}\pi^{0}}_{-}(t)].
\nonumber\\
\end{eqnarray}
The $K_{l3}$ matrix element for neutral kaon can be obtained when we replace $F^{+}_{\mu}$ by
\begin{eqnarray}
F^{-}_{\mu}(p^{\prime},p)  &&\hspace{-0.1cm} = \hspace{0.1cm}<\pi^{-}(p^{\prime})|{\bar s} \gamma_{\mu} u(0)|K^{0}(p)>
\nonumber\\&&\hspace{-0.1cm}
= \frac{1}{\sqrt{2}}[ (p+p^{\prime})_{\mu} f^{K^{0}\pi^{-}}_{+}(t)+(p-p^{\prime})_{\mu}f^{K^{0}\pi^{-}}_{-}(t)].
\nonumber\\
\end{eqnarray}
The four $K_{l3}$ form factors $f^{K^{+}\pi^{0}}_{\pm}(t)$ and $f^{K^{0}\pi^{-}}_{\pm}(t)$ depend on
the four momentum squared which is transferred to the leptons
\begin{equation}
t = (p^{\prime}-p)^{2} = (p_{l}+p_{\nu})^2.
\end{equation}
In the isospin limit $f^{K^{+}\pi^{0}}_{\pm}(t) = f^{K^{0}\pi^{-}}_{\pm}(t)$. In this article
the matrix element in infinite volume is obtained for the neutral koan decay.
The scalar form factor which describes the S-wave projection of the matrix element is defined as
\begin{equation}
\label{scalarform}
f_{0}(t) = f_{+}(t)+ \frac{t}{m_{K}^2-m_{\pi}^2} f_{-}(t).
\end{equation}
At the point of zero recoil, the general decomposition of the vector current matrix element becomes
\begin{eqnarray}
\label{recoil}
<\pi^{-}(p^{\prime})|{\bar s} \gamma_{\mu} u(0)|K^{0}(p)>_{t_{m}}&&\hspace{-0.1cm}  =
\nonumber\\&&\hspace{-4.55cm}
\frac{1}{\sqrt{2}}[ (m_{K}+m_{\pi}) f^{K^{0}\pi^{-}}_{+}(t_{m})
+(m_{K}-m_{\pi})f^{K^{0}\pi^{-}}_{-}(t_{m})]\delta_{\mu 0} \,,
\nonumber\\
\end{eqnarray}
where $t_{m} = (m_{K}-m_{\pi})^{2}$ is the value of momentum transfer at the special point in
the kinematical region.
Given the definition of the scalar form factor in Eq.~(\ref{scalarform}) and the matrix element
in Eq.~(\ref{recoil}) we find the following relation at the maximum momentum transfer
\begin{eqnarray}
\label{scalarform2}
<\pi^{-}(p^{\prime})|{\bar s} \gamma_{\mu} u(0)|K^{0}(p)>_{t_{m}} = \frac{m_{K}+m_{\pi}}{\sqrt{2}} f_{0}(t_{m}) \delta_{\mu 0}.
\nonumber\\
\end{eqnarray}

\section{Analytical results in infinite volume}
\label{loopresult}
\begin{figure}
\begin{center}
\includegraphics[width=0.8\textwidth]{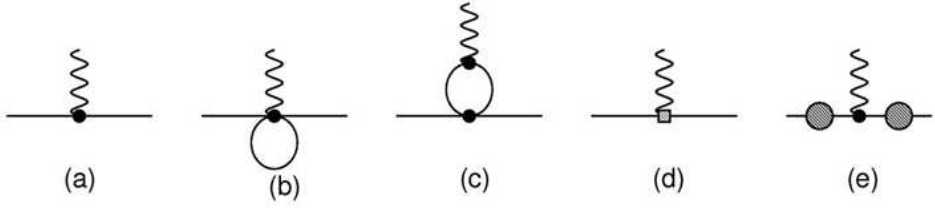}
\caption{Feynman diagrams contributing to the form factors. Diagram (a) is of ${\cal O}(p^2)$ with a vertex
from the ${\cal O}(p^2)$ Lagrangian. The rest of the diagrams are of ${\cal O}{(p^4)}$. Diagram (d) takes its
vertex from the ${\cal O}(p^4)$ Lagrangian and diagram (e) is the wave function corrections to the external legs.
The thick lines define meson propagations and wavy lines indicate the gauge boson.}
\label{diagrams}
\end{center}
\end{figure}

In this section we recalculate the $K_{l3}$ vector form factors at order ${\cal O}(p^2)$ and ${\cal O}(p^4)$
in the isospin limit. On top of this, we also obtain a new expression at one loop order for strangeness changing
matrix element in a tensor form which is useful for finite volume calculations to be discussed in Sec.\,\ref{finiteresult}
and Sec.\,\ref{numerical result}.
In the next subsection we introduce the matrix element in its new form and recall the old results.
Feynman diagrams up to order $p^{4}$ in chiral order needed for the interested process are provided by Fig.~\ref{diagrams}.
\vspace{.5cm}

\subsection{${\cal O}(p^2)$ and ${\cal O}(p^4)$ }
In the isospin limit the vector form factors and scalar form factor at leading order are
\begin{eqnarray}
f^{2}_{+}(t) = f^{2}_{0}(t) = 1 \,,  &&\hspace{.5cm} f^{2}_{-}(t) = 0.
\end{eqnarray}
The superscribe indicates the chiral order. These results are obtained by calculating the tree
Feynman diagram in Fig.~\ref{diagrams}(a).
At the next order there are four diagrams contributing to the matrix element as depicted in Fig.~\ref{diagrams}(b-e).
Without using the tensor simplification relations for the tensor integrals, we evaluate all the diagrams and
sum of the four diagrams read the following expression
\begin{eqnarray}
F(p,p^{\prime}).\epsilon &&\hspace{-0.1cm} = \frac{1}{F_{\pi}^2} \Big [ 2 q^{2} L_{9} + [\frac{3}{8}A(m_{\pi}^2)+\frac{3}{8}A(m_{\eta}^2)
\nonumber\\&&\hspace{-0.1cm}
+\frac{3}{4}A(m_{K}^2)]r.\epsilon
-[\frac{3}{2}B_{\mu \nu}(m_{\pi}^2,m_{K}^2,q^2)
\nonumber\\&&\hspace{-0.1cm}
+\frac{3}{2} B_{\mu \nu} (m_{K}^2,m_{\eta}^2,q^2)]r^{\nu} \epsilon^{\mu}
+[ -2(m_{K}^2 - m_{\pi}^2) L_{9}
\nonumber\\&&\hspace{-0.1cm}
+4(m_{K}^2 - m_{\pi}^2) L_{5} + \frac{1}{2} A(m_{\eta}^2)
-\frac{5}{12}A(m_{\pi}^2)
\nonumber\\&&\hspace{-0.1cm}
+\frac{7}{12} A(m_{K}^2)] q.\epsilon +B(m_{\pi}^2,m_{K}^2,q^2)(\frac{5}{12}q^2
\nonumber\\&&\hspace{-0.1cm}
-\frac{5}{12} m_{K}^2
-\frac{1}{12}m_{\pi}^2) q.\epsilon
+B(m_{K}^2,m_{\eta}^2,q^2)(\frac{1}{4}q^2
\nonumber\\&&\hspace{-0.1cm}
-\frac{7}{12} m_{K}^2
+\frac{1}{12}m_{\pi}^2) q.\epsilon
-[\frac{5}{6}B_{\mu\nu}(m_{\pi}^2,m_{K}^2,q^2)
\nonumber\\&&\hspace{-0.1cm}
+\frac{1}{2}B_{\mu\nu}(m_{K}^2,m_{\eta}^2,q^2)]q^{\mu} \epsilon^{\nu}
+B_{\mu}(m_{\pi}^2,m_{K}^2,q^2)\times
\nonumber\\&&\hspace{-0.1cm}
[\frac{3}{4}(p+p^{\prime})^{\mu} q.\epsilon+\frac{5}{12}q^{\mu} q.\epsilon
+\frac{5}{6}m_{K}^2 \epsilon^{\mu}
+\frac{1}{6} m_{\pi}^2 \epsilon^{\mu}
\nonumber\\&&\hspace{-0.1cm}
-\frac{5}{6} q^{2} \epsilon^{\mu}]
+B_{\mu}(m_{K}^2,m_{\eta}^2,q^2)
[\frac{3}{4}(p+p^{\prime})^{\mu} q.\epsilon
\nonumber\\&&\hspace{-0.1cm}
+\frac{1}{4}q^{\mu} q.\epsilon
+\frac{7}{6}m_{K}^2 \epsilon^{\mu}-\frac{1}{6} m_{\pi}^2 \epsilon^{\mu}
-\frac{1}{2} q^{2}  \epsilon^{\mu}] \Big ],
\end{eqnarray}
with
\begin{eqnarray}
r = p^{\prime}+p \,,  &&\hspace{.5cm}  q = p^{\prime} -p\,,
\end{eqnarray}
and $\epsilon$ is the polarization four vector of the gauge boson.
The integral functions $A, B, B_{\mu}$ and $B_{\mu \nu}$ are introduced
in the Appendix. In infinite volume, it is possible to use the tensor simplification
relations which are provided in the Appendix, to reduce the tensor integrals in terms of scalar functions.
The one loop integrals can be evaluated by applying the dimensional regularization scheme.
The infinities arising from loop integrals can be canceled by redefining (in a renormalization
scheme as discussed in \cite{GL1}) the LEC's in terms of renormalized LEC's
and subtracted parts containing infinities.
After doing the renormalization, we ensure the cancellation of infinities and
recover the old results \cite{Leutwyler:1984}
\begin{eqnarray}
f_+^{K^0\pi^-(4)}(t) &&\hspace{-0.1cm}  =
\frac{1}{F_\pi^2}\Big[ 2\,L^r_{9}t+\frac{3}{8}\,\overline{A}(m_{\pi}^2)+\frac{3}{4}\,\overline{A}(m_{K}^2)
\nonumber\\&&\hspace{-0.1cm}
+\frac{3}{8}\,\overline{A}(m_{\eta}^2)
-\frac{3}{2}\,\overline{B}_{22}(m_{\pi}^2,m_{K}^2,t)
\nonumber\\&&\hspace{-0.1cm}
-\frac{3}{2}\,\overline{B}_{22}(m_{K}^2,m_{\eta}^2,t)\Big]\,,
\end{eqnarray}
\begin{eqnarray}
f_-^{K^0\pi^-(4)}(t) &&\hspace{-0.1cm} =
\frac{1}{F_\pi^2}\,\Big[-2\,m_{K}^2\,L^r_{9}+4\,m_{K}^2\,L^r_{5}-4\,m_{\pi}^2\,L^r_{5}
\nonumber\\&&\hspace{-0.1cm}
+2\,m_{\pi}^2\,L^r_{9}-\frac{5}{12}\,\overline{A}(m_{\pi}^2)
       +\frac{7}{12}\,\overline{A}(m_{K}^2)
      \nonumber\\&&\hspace{-0.1cm}
       +\frac{1}{2}\,\overline{A}(m_{\eta}^2)+\overline{B}(m_{\pi}^2,m_{K}^2,t)\,
       (\frac{5}{12}\,t-\frac{5}{12}\,m_{K}^2
      \nonumber\\&&\hspace{-0.1cm}
       -\frac{1}{12}\,m_{\pi}^2)+\overline{B}(m_{K}^2,m_{\eta}^2,t)\,(\frac{1}{4}\,t
       -\frac{7}{12}\,m_{K}^2
      \nonumber\\&&\hspace{-0.1cm}
       +\frac{1}{12}\,m_{\pi}^2 )+\,\overline{B}_{1}(m_{\pi}^2,m_{K}^2,t)\,(-\frac{5}{12}\,t+\frac{19}{12}\,m_{K}^2
      \nonumber\\&&\hspace{-0.1cm}
       -\frac{7}{12}\,m_{\pi}^2)+\,\overline{B}_{1}(m_{K}^2,m_{\eta}^2,t)\,(-\frac{1}{4}\,t+\frac{23}{12}\,m_{K}^2
      \nonumber\\&&\hspace{-0.1cm}
       -\frac{11}{12}\,m_{\pi}^2)+\,\overline{B}_{21}(m_{\pi}^2,m_{K}^2,t)\,(-\frac{5}{6}\,t-\frac{3}{2}\,m_{K}^2
       \nonumber\\&&\hspace{-0.1cm}
       +\frac{3}{2}\,m_{\pi}^2)+\overline{B}_{21}(m_{K}^2,m_{\eta}^2,t)\,(-\frac{1}{2}\,t-\frac{3}{2}\,m_{K}^2
       \nonumber\\&&\hspace{-0.1cm}
       +\frac{3}{2}\,m_{\pi}^2)-\frac{5}{6}\,\overline{B}_{22}(m_{\pi}^2,m_{K}^2,t)
       \nonumber\\&&\hspace{-0.1cm}
       -\frac{1}{2}\,\overline{B}_{22}(m_{K}^2,m_{\eta}^2,t) \Big].
\end{eqnarray}

\section{Finite volume calculations}
\label{finiteresult}
When we enclose a system in a finite box, Lorentz symmetry is no longer respected. This can make it impossible
to define without ambiguity vector form factors for such a system. It also turns out impossible to define
vector form factors at the point of zero recoil energy in finite volume because it is always possible
to reshuffle terms in the matrix element in such a way to obtain different values for $f_{+}$ and $f_{-}$ according
to their definitions in Eq.(\ref{recoil}). However, we see from Eq.(\ref{scalarform2})
that the time component of the matrix element at the maximum momentum transfer, $t_{m}$, is proportional to the scalar
form factor and therefore the evaluation of the time component of the matrix element
in finite volume can give us the scalar form factor at $t_{m}$.
The one-loop Feynman integrals which contribute to the matrix element of the process are
evaluated in this section. We expect finite volume effects arising from the simultaneous propagation
of kaon and eta particles to be much smaller than the same effects when pion and kaon
propagate simultaneously in the loop. We demonstrate this claim quantitatively in the next
subsection.

\subsection{Scalar one-loop integrals}
We work out in this subsection the one-loop scalar integrals in detail.
In what follows we define for a generic function, F, $\Delta F = F_{V} - F_{\infty}$.
Subscripts $V$ and $\infty$ indicate integration in finite
and infinite volume, respectively. In finite volume, as we mentioned before,
momentum $\vec q$ gets discrete values, $\vec q = \frac{2\pi}{L} \vec{n}$,
but we obtain our formula for the case with infinite time extension.
We begin with the tadpole integral
\begin{eqnarray}
\label{afunc}
A_{V} (m) &&\hspace{-0.1cm} = -\frac{i}{L^3} \sum_{\vec{q}} \int \frac{dq_0}{2\pi} \frac{1}{q^2-m^2}
\nonumber\\&&\hspace{-0.1cm}
          =  A_{\infty} - i \int \frac{dq_0}{2\pi}\sum_{\vec{n}\neq 0}\int \frac{d^{3}\vec{q}}{(2\pi)^3} \frac{e^{iL\vec{q}.\vec{n}}}{q^2-m^2}.
\end{eqnarray}
To get the second equality we have used the Poisson summation formula
\begin{eqnarray}
\frac{1}{L^3} \sum_{\vec{q}=\frac{2\pi}{L} \vec{n}} f(\vec{q}~^2) = \int \frac{d^3q}{(2\pi)^3} f(\vec{q}~)
+\sum_{\vec{n}\neq 0} \int \frac{d^3q}{(2\pi)^3} f{(\vec q~)} e^{iL\vec{q}.\vec{n}}.
\end{eqnarray}
By evaluating the contour integral over $q_{0}$ in Eq.(\ref{afunc}) we arrive at
\begin{eqnarray}
\label{A1}
A_{V}(m) = A_{\infty}(m) - \sum_{\vec{n}\neq 0}\int \frac{d^3\vec{q}}{(2\pi)^{3}} \frac{e^{iL\vec{q}.\vec{n}}}{2\sqrt{\vec{q}^2+m^2}}.
\end{eqnarray}
We finally take the final integral and obtain
\begin{eqnarray}
\Delta A = - \frac{m}{4 \pi^2 L} \sum_{\vec{n}\neq \vec{0}} \frac{1}{\vert \vec{n} \vert} m(k) K_{1}(m L \vert \vec{n}\vert),
\end{eqnarray}
where function $K_{1}$ is the modified Bessel function of order one. m(k) accounts
for the number of ways that the equality $\vec{n}^2 = n_{1}^2+n_{2}^2+n_{3}^2$
is satisfied for positive and negative integer numbers
for $n_{1}$, $n_{2}$ and $n_{3}$. This result agrees with that in \cite{Bijnens-Ghorbani2006}.

\begin{figure}
\begin{center}
\includegraphics[width=0.8\textwidth]{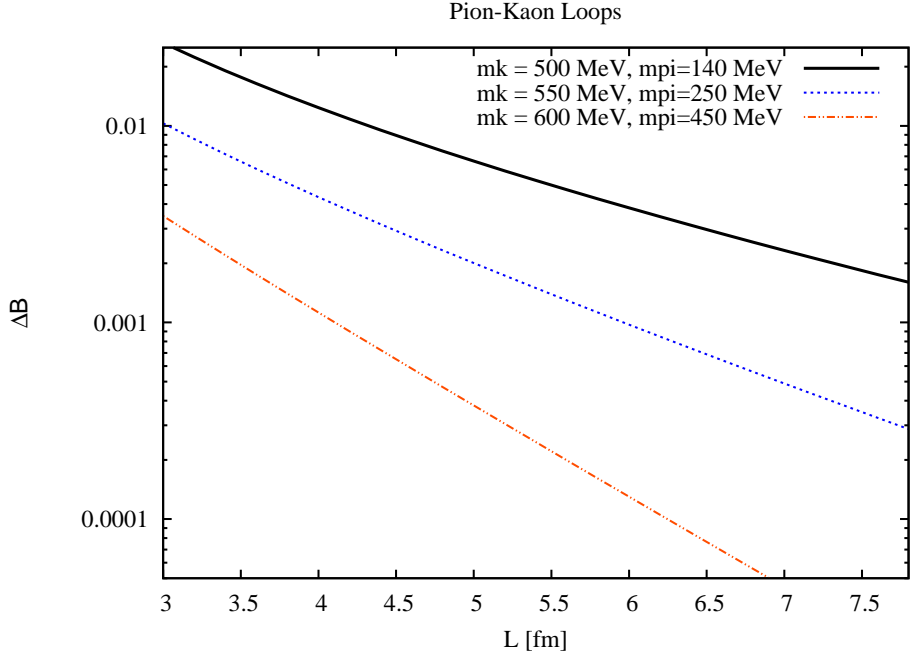}
\caption{$\pi$-K loops in finite volume for three different values of the pion mass and kaon mass.}
\label{Bfunction1}
\end{center}
\end{figure}

\begin{figure}
\begin{center}
\includegraphics[width=0.8\textwidth]{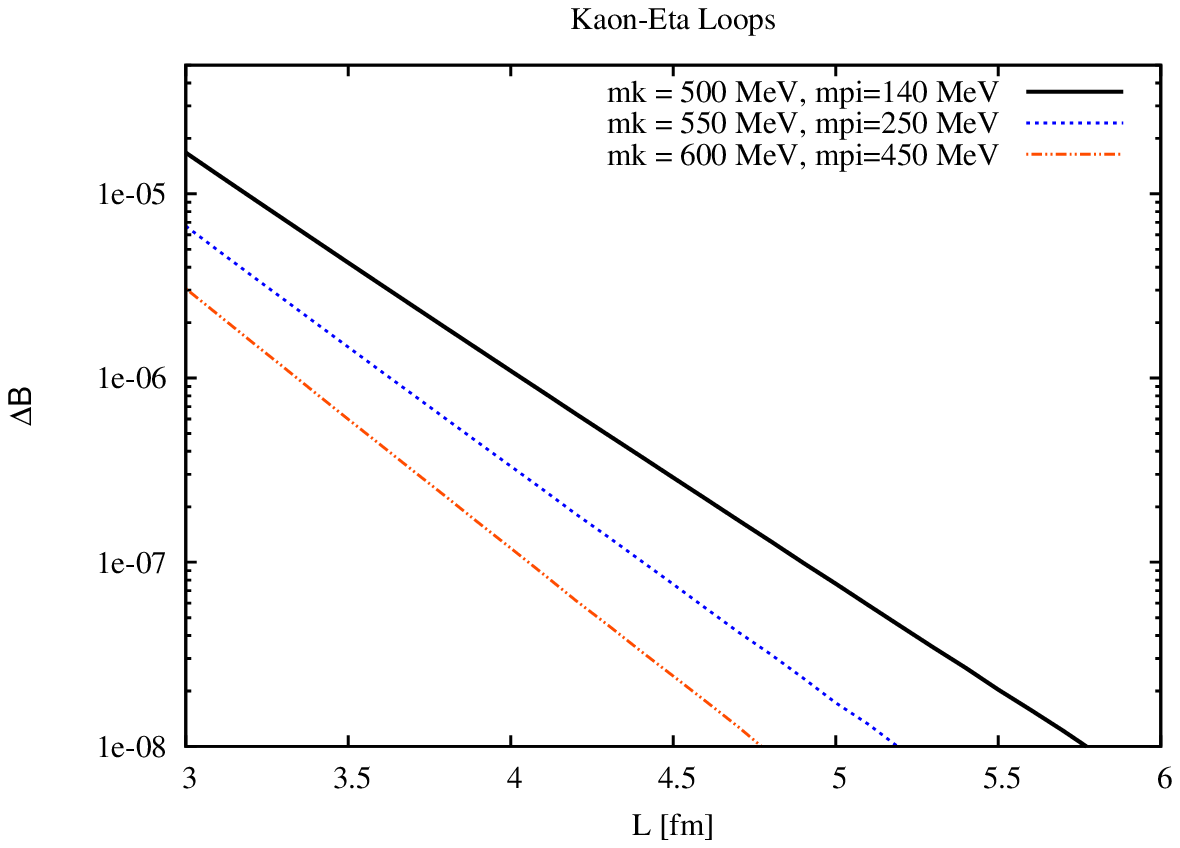}
\caption{K-$\eta$ loops in finite volume for three different values of the pion mass and kaon mass }
\label{Bfunction2}
\end{center}
\end{figure}

The second Feynman integral we need to consider here, involves the propagation of two particles with different masses.
At the momentum transfer, $t_{m} = (M_{K}-M_{\pi})^2$, we first perform the integral for particles with masses $M=M_{K}$
and $m=M_{\pi}$ in the propagators
\begin{eqnarray}
\label{bfunc}
B_{V}&&\hspace{-0.1cm}= -\frac{i}{L^3} \sum_{\vec{q}} \int \frac{dq_0}{2\pi} \frac{1}{q^2-m^2} \frac{1}{(q+p)^2-M^2}
                = B_{\infty}
\nonumber\\&&\hspace{-0.2cm}
                -\int\frac{dq_0}{2\pi}\sum_{\vec{n}\neq 0}\int \frac{d^{3}\vec{q}}{(2\pi)^3}
                \frac{i e^{iL\vec{q}.\vec{n}}}{(q^2-m^2)((q+p)^2-M^2)}\,,
                \nonumber \\
\end{eqnarray}
where the Poisson summation formula is employed to get the second line.
For a simplifying assumption we take the external momentum completely in the time direction as $p = (p_{0},\vec{0})$ and
we proceed by exploiting the Feynman parameter formula to get
\begin{eqnarray}
\label{bfunc2}
B_{V}= B_{\infty}
- i \int_{0}^{1}dx \int \frac{dq_0}{2\pi} \sum_{\vec{n}\neq 0}
\int \frac{d^{3}\vec{q}}{(2\pi)^3}
\frac{e^{iL\vec{q}.\vec{n}}}{[q^2+x(1-x)p_{0}^2-xm^2-(1-x)M^2]^{2}}\,.
\nonumber \\&&
\end{eqnarray}
We first perform a contour integration over $q_{0}$ and then plug in $p_{0}^2 = t_{m}$ to find
\begin{eqnarray}
B_{V} = B_{\infty}
+\frac{1}{4} \sum_{\vec{n}\neq 0}\int \frac{d^{3}\vec{q}}{(2\pi)^3}
e^{iL\vec{q}.\vec{n}}
\int_{0}^{1} \frac{dx}{[\vec{q}^2+(x m+(x-1)M)^2]^{3/2}}.
\end{eqnarray}
Taking the integral over $x$ gives
\begin{eqnarray}
\label{B1}
B_{V} = B_{\infty}+\frac{1}{4(M-m)} \sum_{\vec{n}\neq 0}
\int \frac{d^{3}\vec{q}}{(2\pi)^3}
e^{iL\vec{q}.\vec{n}}
(\frac{M}{\vec{q}^2 \sqrt{\vec{q}^2+M^2}}-\frac{m}{\vec{q}^2 \sqrt{\vec{q}^2+m^2}}).
\end{eqnarray}
The second piece contains a three dimensional integral where integration over the angular part leaves us with
\begin{eqnarray}
B_{V} = B_{\infty}
-\frac{i}{(2\pi)^2 4(M-m)L} \sum_{\vec{n}\neq 0} \frac{1}{\vert \vec{n} \vert}
\int_{-\infty}^{+\infty} d \vert \vec{q} \vert \hspace{.1cm}
(\frac{M e^{iL \vert \vec{q} \vert  \vert \vec{n} \vert }}{\vert \vec{q} \vert \sqrt{{\vert \vec{q} \vert}^2+M^2}}
-\frac{m e^{iL \vert \vec{q} \vert  \vert \vec{n} \vert }}{\vert \vec{q} \vert \sqrt{{\vert \vec{q} \vert}^2+m^2}} ).
\nonumber \\&&
\end{eqnarray}
To carry out the last integration we make use of the convolution technique. After taking the integral we obtain
\begin{eqnarray}
\Delta B =
\frac{1}{16 \pi (M-m)} \sum_{\vec{n} \neq 0}
M [L_{-1}(M L \vert \vec{n} \vert) K_{0}(M L \vert \vec{n} \vert)
+L_{0}(M L \vert \vec{n} \vert) K_{1}(M L \vert \vec{n} \vert)]
\nonumber\\&&\hspace{-10.cm}
-m [L_{-1}(m L \vert \vec{n} \vert) K_{0}(m L \vert \vec{n} \vert)
+L_{0}(m L \vert \vec{n} \vert) K_{1}(m L \vert \vec{n} \vert)].
\nonumber\\&&
\end{eqnarray}
$L_{\alpha}$ and $K_{\alpha}$ are respectively modified Struve and modified Bessel functions of order $\alpha$.
In Fig.~\ref{Bfunction1}, the function $\Delta B$ is plotted versus $L$ for three different values of the pion mass
while we change the kaon mass correspondingly. The results in Fig.~\ref{Bfunction1} show a reasonable trend.
In addition, we need to perform another one-loop integral for propagating particles with masses $M_{\eta}$
and $M_{K}$. It turned out that one cannot find a formula in closed form in this case. The final result
for the $K-\eta$ loops reads
\begin{eqnarray}
\Delta B =
\frac{1}{8 \pi^2} \sum_{\vec{n} \neq 0}
\frac{1}{L \vert \vec{n} \vert} \Big( \frac{H_{1}+H_{2}}{H_{1}}
\int_{0}^{\infty}
\frac{Sin(L \vert \vec{n} \vert q) q dq}{(\vec{q}^2+H_{3})\sqrt{(H_{1}+H_{2})^2+H_{3}+\vec{q}^2}}
\nonumber\\&&\hspace{-8.5cm}
-\frac{H_{2}}{H_{1}}\int_{0}^{\infty}
\frac{Sin(L \vert \vec{n} \vert q) q dq}{(\vec{q}^2+H_{3})\sqrt{H_{2}^2+H_{3}+\vec{q}^2}} \Big) \,,
\nonumber\\&&
\end{eqnarray}
where
\begin{eqnarray}
H_{1} = M_{K} - M_{\pi} \,,
\nonumber\\&&\hspace{-3.7cm}
H_{2}= \frac{M_{K}^2 - M_{\eta}^2 - (M_{K}-M_{\pi})^2}{2(M_{K}-M_{\pi})}\,,
\nonumber\\&&\hspace{-3.7cm}
H_{3} = M_{\eta}^2 - H_{2}^2.
\end{eqnarray}
We solve the integral above numerically and plot the function $\Delta B$ versus $L$ as depicted
in Fig.~\ref{Bfunction2}. By comparing the results in Fig.~\ref{Bfunction1} and Fig.~\ref{Bfunction2}, we conclude that
the $K-\eta$ loops are suppressed with respect to the $\pi-K$ loops.

\subsection{Tensor one-loop integrals}
The tensor integral $B^{\mu}$ defined in the Appendix has only one component, i.e. $B^{0}$, needed to be evaluated
in the finite volume at the momentum transfer $t_{m}$. We thus consider the integral
\begin{eqnarray}
B_{V}^{0} = -\frac{i}{L^3} \int \frac{dq_{0}}{2\pi} \sum_{\vec{q}} \frac{q_{0}}{(q^2-m^2)((q+p)^2-M^2)}
= B^{0}_{\infty} + \int \frac{dq_{0}}{2\pi} \sum_{\vec{n}\neq 0} \int \frac{d^3 \vec{q}}{(2\pi)^3}
\nonumber\\&&\hspace{-8.5cm}
\times \frac{q_{0} \hspace{+.1cm} e^{iL\vec{q}.\vec{n}}}{(q^2-m^2)((q+p)^2-M^2)}.
\end{eqnarray}
The second term which we call $\Delta B^{0}$ can be written in terms of
the functions which we obtained already. After doing some manipulations on the integrand, we obtain
\begin{eqnarray}
\Delta B^{0} = \frac{1}{2(M-m)} [\Delta A(m)-\Delta A(M)
+2m(M-m)\Delta B].
\end{eqnarray}

The last integral we need to calculate in the finite volume is the $B^{00}$ component of the tensor integral $B^{\mu \nu}$.
Again we apply the Poisson summation formula to arrive at the second line for the integral below
\begin{eqnarray}
B_{V}^{00}&&\hspace{-0.1cm} = -\frac{i}{L^3} \int \frac{dq_{0}}{2\pi} \sum_{\vec{q}} \frac{q_{0}^{2}}{(q^2-m^2)((q+p)^2-M^2)}
\nonumber\\&&\hspace{-0.1cm}
= B^{00}_{\infty} - i \int \frac{dq_{0}}{2\pi} \sum_{\vec{n}\neq 0} \int \frac{d^3 \vec{q}}{(2\pi)^3}
\frac{q_{0}^2 \hspace{+.1cm} e^{iL\vec{q}.\vec{n}}}{(q^2-m^2)((q+p)^2-M^2)}.
\end{eqnarray}
We integrate over $q_{0}$ following the same argument sketched before and find
\begin{eqnarray}
\Delta B^{00}&&\hspace{-0.1cm} = -\int \frac{d^3\vec{q}}{(2\pi)^3} \frac{e^{iL\vec{q}.\vec{n}}}{2\sqrt{\vec{q}^2+M^2}}
+\frac{1}{4(M-m)} \int \frac{d^3\vec{q}}{(2\pi)^3}
\nonumber\\&&\hspace{.6cm}
\times e^{iL\vec{q}.\vec{n}}
(\frac{M(\vec{q}^2+m^2)}{\vec{q}^2 \sqrt{{\vec{q}}^2+M^2}}-\frac{m(\vec{q}^2+m^2)}{\vec{q}^2 \sqrt{{\vec{q}}^2+m^2}})\,.
\end{eqnarray}
Substituting the results of Eq.(\ref{A1}) and Eq.(\ref{B1}) into the relation above will yield us
\begin{eqnarray}
\Delta B^{00}&&\hspace{-0.1cm} = \Delta A(M) - \frac{1}{2(M-m)} [M\Delta A(M)-m \Delta A(m)]
+ m^2 \Delta B.
\end{eqnarray}

\subsection{An asymptotic formula for the scalar form factor}

It is important to know the behavior of the finite volume
correction of the scalar form factor in asymptotically
large volume. To this end, we provide a formula in which
we have ignored eta loop effects as well as terms with
kaon mass in the exponential and obtain the following result
\begin{eqnarray}
\label{asymptotic}
\Delta f_{0}(t_{m})&&\hspace{-0.1cm}  \simeq \frac{1}{F_{\pi}^2} \Big( \frac{ \sqrt{2m_{\pi}}(m_{K}-19 m_{\pi})}{24(2\pi L)^{3/2}(m_{\pi}+m_{K})}
\nonumber\\&&\hspace{-0.1cm}
+\frac{ \frac{34}{27}m_{\pi}^3-\frac{5}{12} m_{\pi}^2 m_{K}+\frac{1}{4}m_{K}^2 m_{\pi}}{2(\pi L)^{3/2}\sqrt{m_{\pi}} (m_{K}^2-m_{\pi}^2)}
 \Big) \times
 \nonumber\\&&\hspace{-0.1cm}
  \sum_{\vec{n} \neq 0} \frac{e^{-m_{\pi} L \vert \vec{n} \vert}}{{\vert \vec{n} \vert}^{3/2}}.
\end{eqnarray}
Hence, one clearly notices that the volume corrections to the scalar form factor are exponentially suppressed.

\section{Numerical results and conclusions}
\label{numerical result}
\begin{figure}
\begin{center}
\includegraphics[width=0.8\textwidth]{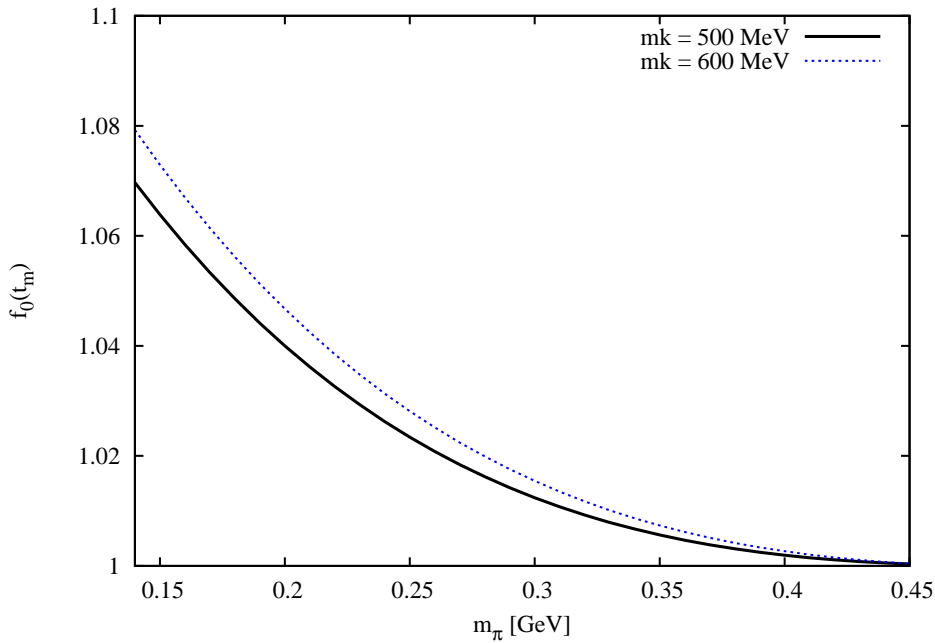}
\caption{We compare the pion mass dependence of the scalar form factor at the endpoint in the infinite volume
for two different values of the kaon mass.}
\label{fscalar}
\end{center}
\end{figure}

It is seen that the value of the scalar form factor at
the endpoint in infinite volume only depends on
one low energy constant i.e., $L_{5}^{r}$, which we
use as input $L_{5}^{r} = 0.97205 \times 10^{-3}$, see
\cite{Amoros2001} for detailed discussion on this LEC.
The subtraction scale is $\mu = 770\,MeV$. There is an ambiguity
on the value we should put in for the pion decay constant
when one calculates a quantity at one loop order. We use
the physical value for the pion decay constant
instead of its lowest order value since the difference
shifts to higher order. The value $F_{\pi} = 0.0924\,GeV$
is chosen as input. The loop integrals in infinite volume
are performed with the renormalization scale
$\mu = 770\,MeV$.
We show in Fig.~\ref{fscalar} the pion mass
dependence of the scalar form factor in infinite volume
at the endpoint for $m_{K} = 500\,MeV$ and $m_{K} = 600\,MeV$.
Moreover, from Fig.~\ref{fscalar}, one can see that in going
away from the physical value of the pion mass to an unphysical
value of about $350\,MeV$ the scalar form factor in the
infinite volume changes about $5$ percent.
In addition, we study the volume dependence of
the form factor at order $p^{4}$ by defining the ratio
\begin{eqnarray}
R_{f_{0}} =\frac{f_{0}^{V}(t_{m}) - f_{0}^{\infty}(t_{m})}{f_{0}^{\infty}(t_{m})}.
\end{eqnarray}

\begin{figure}
\begin{center}
\includegraphics[width=0.8\textwidth]{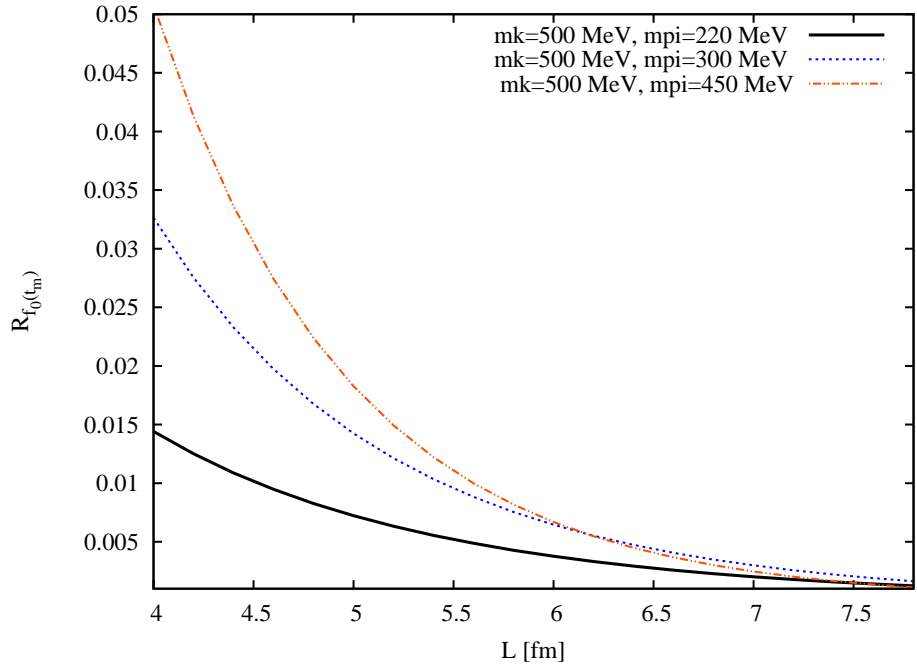}
\caption{The ratio $R_{f_{0}}$ is plotted versus the linear size of the lattice
for pion masses $m_{\pi} = 220\,MeV$,  $m_{\pi} = 300\,MeV$ and $m_{\pi} = 450\,MeV$ for a fixed value of
the kaon mass, $m_{K} = 500\,MeV$.}
\label{ratio-mass}
\end{center}
\end{figure}

The ratio $R_{f_{0}}$ is plotted in Fig.~\ref{ratio-mass}
versus lattice size, L, for three different
values of the pion mass, i.e. $m_{\pi} = 220\,MeV$,
$m_{\pi} = 300\,MeV$ and $m_{\pi} = 450\,MeV$,
with a fixed value for the kaon mass, $m_{K} = 500\,MeV$.
The results in Fig.~\ref{ratio-mass} clearly show that
for lattice sizes smaller than $L= 6.15\,fm$
there is a larger finite size effects for the larger
pion mass while the ratio behaves differently for the
lattice sizes larger than $L = 6.15\,fm$.
We found out that the term with the function $B^{0}$
in our finite volume expression for the
scalar form factor, takes responsibility for finite
volume effects to become large for larger pion mass
at small volumes. The reason is that the function $B^{0}$
in the finite volume does not converge fast
enough to compensate the quadratic growth of the pion mass
in front of $B^{0}$. However, as we see from Eq.~\ref{asymptotic},
the finite volume corrections to the scalar
form factor exhibit a standard behavior at asymptotically
large pion mass, wherein the corrections suppress exponentially.
We therefore stress that the finite volume effects
for the scalar form factor at the endpoint can be sizable
for $L= 4\,fm$ for large pion mass of about $300\,MeV$.

\begin{table}
  \centering
\scriptsize{
   \begin{tabular}{|c|c|c|c|c|c|c|}
\hline
  $m_{\pi} (GeV)$ & $m_{K} (GeV)$  & $m_{\pi}$L & $\Delta f_{0}$ (Only $\pi-K$ loops)& $\Delta f_{0}$ (Including $K-\eta$ loops) & $f^{V}_{0}$ & $f^{V}_{0}$ \cite{Boyle2007a}\\
\hline
   \hline
  0.329   & 0.575  &   4.50   & 0.08459 & 0.08490  & 1.09532  & 1.02143(132) \\
  \hline
  0.416   & 0.604  &   5.69   & 0.11169 & 0.11196  & 1.11360  & 1.00887(89)  \\
  \hline
  0.556   & 0.663  &   7.61   & 0.12323 & 0.12340  & 1.12179  & 1.00192(34)  \\
  \hline
  0.671 & 0.719    &   9.19   & 0.11435 & 0.11446  & 1.11382    & 1.00029(6)   \\
  \hline
\end{tabular}
}
\caption{\label{tabl1} We compare our finding for the scalar form factor in finite volume with those quoted
in \cite{Boyle2007a}, corresponding to the pion masses and kaon masses given in the first and second column.
The linear size of the lattice in these calculations is 2.74~fm.
Finite volume corrections to the scalar form factor without and with K-$\eta$ loop contributions are given in the
forth and fifth column respectively.}
\end{table}

In Table 1. we present the finite volume corrections to
the scalar form factor and the value of the
form factor in finite volume with $L = 2.73\,fm$ for
particular values of the pion mass and kaon mass as used in
\cite{Boyle2007a}. Given the systematic errors in the
scalar form factor quoted in \cite{Boyle2007a}
our results in Table.~\ref{tabl1} show that the lattice
data is precise enough for heavier pion masses
to take into account the $\eta$ contributions at finite volume.
However, our results for the finite volume correction to the scalar
form factor at finite volume appear to be large so that the
resulting scalar form factor in finite volume
is larger than those quoted in \cite{Boyle2007a}
while for light pion masses the discrepancy
becomes smaller.
This is expected from Fig.~\ref{ratio-mass}, since
our calculated finite volume corrections become large
for large pion masses while being in the {\it p-regime}, and
therefore this is the question of how much SU(3) ChPT
is reliable for larger pion masses in the process studied
in this article.
Moreover, we have also evaluated the
finite size effects for a smaller lattice size with
$L = 1.83\,fm$ and find that the finite volume effects
here is about 40 percent which exceed the almost
10 percent finite volume corrections obtained for a lattice
with $L = 2.73\,fm$. The difference between the estimated
finite size effects from ChPT for $L = 2.73\,fm$ and
$L = 1.83\,fm$ is quite larger than the amount
of difference as quoted in Table III
found in \cite{Boyle2007a}. This is again
due to the fact the used pion masses in our
ChPT calculations for the scalar form factor are large.
\\

Further work in this direction is to extend
our results to non-vanishing spatial momentum since
lattice practitioners use these values to interpolate
to zero momentum transfer. Moreover, our study can
be extended to the case of twisted boundary condition.

\vspace{1cm}
\section*{Acknowledgments}
We would like to thank Johan Bijnens and Gilberto Colangelo for useful comments.
We wish to thank the Institute for research in fundamental sciences (IPM)
for their hospitality while this research was carried out.
\section{Appendix}
We introduce the one loop Feynman integrals as follows
\begin{eqnarray}
A(m^2) =  \frac{1}{i} \int \frac{d^dq}{(2\pi)^d} \frac{1}{q^2-m^2},
\end{eqnarray}
\begin{eqnarray}
B(m^2,M^2,p^2)&&\hspace{-0.1cm} =  \frac{1}{i} \int \frac{d^dq}{(2\pi)^d}
\nonumber\\&&\hspace{.5cm}
\times \frac{1}{(q^2-m^2)((q+p)^2-M^2)},
\end{eqnarray}
\begin{eqnarray}
B_{\mu}(m^2,M^2,p^2)&&\hspace{-0.1cm} =  \frac{1}{i} \int \frac{d^dq}{(2\pi)^d}
\nonumber\\&&\hspace{.5cm}
\times \frac{p_{\mu}}{(q^2-m^2)((q+p)^2-M^2)},
\end{eqnarray}
\begin{eqnarray}
B_{\mu \nu}(m^2,M^2,p^2)&&\hspace{-0.1cm} =  \frac{1}{i} \int \frac{d^dq}{(2\pi)^d}
\nonumber\\&&\hspace{.5cm}
\times \frac{p_{\mu} p_{\nu}}{(q^2-m^2)((q+p)^2-M^2)}.
\end{eqnarray}
Invoking lorentz symmetry, the tensor integrals above can be written in terms of scalar functions
\begin{eqnarray}
B_{\mu}(m^2,M^2,p^2) =  p_{\mu} B_{1}(m^2,M^2,p^2) ,
\end{eqnarray}
\begin{eqnarray}
B_{\mu \nu}(m^2,M^2,p^2)&&\hspace{-0.1cm} =  p_{\mu} p_{\nu} B_{12}(m^2,M^2,p^2)
\nonumber\\&&\hspace{-.1cm}
+g_{\mu \nu} B_{22}(m^2,M^2,p^2).
\end{eqnarray}

%

\end{document}